\documentclass{aa}

\usepackage{natbib}
\usepackage{graphicx}
\usepackage{color}
\begin{document}

\title{Formation of a tiny flux rope in the center of an active region driven by magnetic flux emergence, convergence, and cancellation}

\author{Ruisheng Zheng\inst{1}, Yao Chen\inst{1}, Bing Wang\inst{1}, Hongqiang Song\inst{1}, and Wenda Cao\inst{2,3}}

\institute{Shandong Key Laboratory of Optical Astronomy and Solar-Terrestrial Environment, School of Space Science and Physics, Institute of Space Sciences, Shandong University, Weihai, Shandong, 264209, China \\
              \email{ruishengzheng@sdu.edu.cn}\\
              \and
              Big Bear Solar Observatory, New Jersey Institute of Technology, 40386 North Shore Lane, Big Bear City, CA 92314-9672, USA \\
              \and
              Center for Solar-Terrestrial Research, New Jersey Institute of Technology, University Heights, Newark, NJ 07102-1982, USA}

\date{Received / Accepted }
\titlerunning{Formation and eruption of tiny flux rope}
    \authorrunning{Zheng et al.}

\abstract
% context
{}
% aims
{Flux ropes are generally believed to be core structures of solar eruptions that are significant for the space weather, but their formation mechanism remains intensely debated. We report on the formation of a tiny flux rope beneath clusters of active region loops on 2018 August 24.
}
% methods
{Combining the high-quality multiwavelength observations from multiple instruments, we studied the event in detail in the photosphere, chromosphere, and corona.
}
%results
{In the source region, the continual emergence of two positive polarities (P1 and P2) that appeared as two pores (A and B)is unambiguous. Interestingly, P2 and Pore B slowly approached P1 and Pore A, implying a magnetic flux convergence. During the emergence and convergence, P1 and P2 successively interacted with a minor negative polarity (N3) that emerged, which led to a continuous magnetic flux cancellation. As a result, the overlying loops became much sheared and finally evolved into a tiny twisted flux rope that was evidenced by a transient inverse S-shaped sigmoid, the twisted filament threads with blueshift and redshift signatures, and a hot channel.
}
%conclusions
{All the results show that the formation of the tiny flux rope in the center of the active region was closely associated with the continuous magnetic flux emergence, convergence, and cancellation in the photosphere. Hence, we suggest that the magnetic flux emergence, convergence, and cancellation are crucial for the formation of the tiny flux rope.
}


\maketitle
\section{Introduction}
Magnetic flux ropes, consisting of twisted field lines that globally wrap around the central axis, are considered as core structures of coronal mass ejections (CMEs). In the corona, flux ropes have various manifestations, such as sigmoids, filaments, filament channels, coronal cavities, and hot channels, depending on the magnetic environment of the structure (Cheng et al. 2017). In interplanetary space, flux ropes can be identified as magnetic clouds at 1 AU, and their expansions facilitate the propagations of CMEs and the occurrence of hazardous space weather. Hence, flux ropes are essential to understanding solar eruptions and space weather. The most important issue of flux ropes is the formation and eruption mechanism, which has been intensely debated.

There are two main categories for the initiation models of the flux rope eruption. The first is the magnetic reconnection category, including the breakout model (Antiochos et al. 1999), flux emergence model (Chen \& Shibata 2000), and tether-cutting model (Moore et al. 2001). Another category suggests that the flux rope eruption is triggered by the catastrophic model (Forbes \& Isenberg 1991), kink instability (T{\"o}r{\"o}k et al. 2004), or torus instability (Kliem \& T{\"o}r{\"o}k 2006) in ideal magnetohydrodynamic (MHD) models.

For the formation mechanisms, two possibilities have been proposed. One is the bodily emergence from the convection zone into the corona (Lites et al. 1995; Fan 2001; Manchester et al. 2004; Leak et al. 2013), and another is the direct formation in the corona prior to or during the eruption by magnetic reconnections (Cheng et al. 2011; Zhang et al. 2012; Song et al. 2014; Wang et al. 2017; Threlfall et al. 2018). The magnetic reconnections are always driven by photospheric motions, such as sunspot rotation, shear motion, flux cancellation, and flux convergence (van Ballegooijen \& Martens 1989; Fan 2009; Green et al. 2011; Savcheva et al. 2012; Xia et al. 2014; Kumar et al. 2015; Yan et al. 2016; Zhao et al. 2017). Moreover, the flux rope can also be formed in confined eruptions (Patsourakos et al. 2013; Liu et al. 2018).

In this work, we present the formation of a tiny flux rope in the center of an active region (AR), which was intimately associated with the magnetic flux emergence, convergence, and cancellation in the photosphere. This event sheds light into the formation mechanism of flux ropes.

\section{Observations and data analysis}
%%%%%%%%%%%%%%%%%%%%%%%%%%%%%%%%%%%%%%%%%%%%%%%%%%%%%%%%%%%%%%%%%%%%%%%%%%%%%%%%%%%%%%%%%%%%%%%%%%%%%%%%%%%%%%%%%%%%%
\begin{figure*}
\centering
\includegraphics[width=8cm]{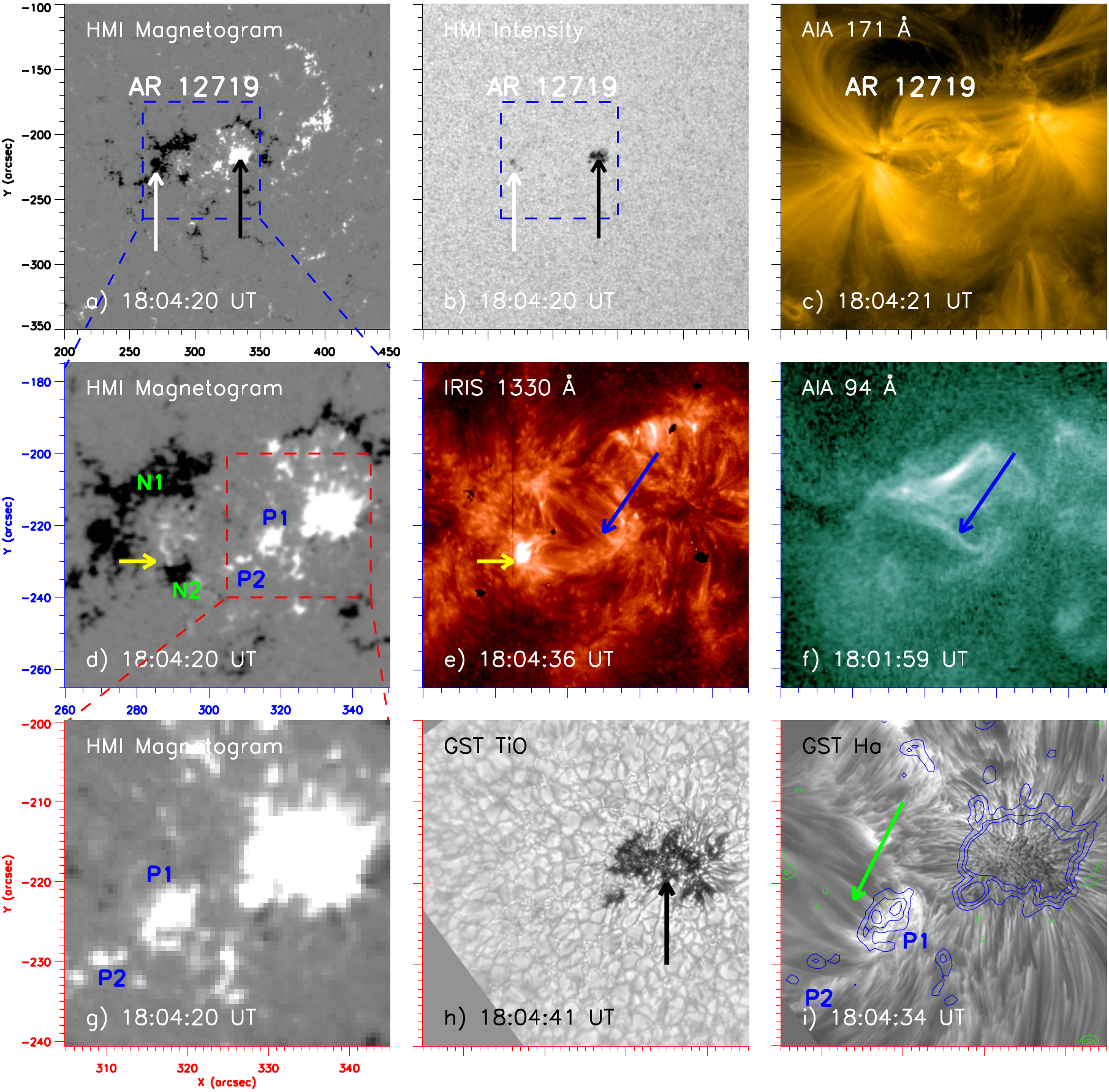}
\caption{Overview of AR 12719 and magnetic polarities (N1-N2 and P1-P2) before the eruption in HMI magnetogram and intensity, GST TiO and H$\alpha$, IRIS 1330~{\AA}, and AIA 171 and 94~{\AA}. The white and black arrows indicate the sunspots, and the yellow arrows show a magnetic cancellation site around N2. The blue and green arrows indicate the coronal loops (L1) and the filament threads of interest, respectively. The dashed blue and red boxes represent the field of view of the middle and bottom panels, respectively. Contours of HMI longitudinal magnetic fields at 18:04:20 UT are superposed on panels (i) with positive (negative) fields in blue (green), and the levels for positive (negative) fields are 200 (50), 400 (100), and 600 (150) gauss.}
\label{f1}
\end{figure*}
%%%%%%%%%%%%%%%%%%%%%%%%%%%%%%%%%%%%%%%%%%%%%%%%%%%%%%%%%%%%%%%%%%%%%%%%%%%%%%%%%%%%%%%%%%%%%%%%%%%%%%%%%%%%%%%%%%%%%
The formation and eruption of a tiny flux rope occurred in NOAA AR 12719 ($\sim$S07W22) on 2018 August 24. The related filament threads and pores were captured well by the high-resolution observations of H$\alpha$ and TiO from the 1.6 m Goode Solar Telescope (GST; Cao et al. 2010) at Big Bear Solar Observatory. The GST employs a high-order adaptive optics system with 308 subapertures and post-facto speckle image reconstruction techniques (W{\"o}ger et al. 2008) to accomplish diffraction-limited imaging of the solar atmosphere. The H$\alpha$ and TiO images have pixel resolutions of 0$\farcs$029 and 0$\farcs$034, respectively, and their cadences are 26 s and 15 s, respectively. The H$\alpha$ observations consist of images at the H$\alpha$ line center and at the line wings of $\pm$0.4, $\pm$0.6, $\pm$0.8, and $\pm$1.0~{\AA}.

We also used full-disk observations from the Helioseismic and Magnetic Imager (HMI; Scherrer et al. 2012) and from the Atmospheric Imaging Assembly (AIA; Lemen et al. 2012) onboard the Solar Dynamics Observatory (SDO; Pesnell et al. 2012). Magnetograms and intensity maps from the HMI, with a cadence of 45 s and pixel scale of 0$\farcs$6, are utilized to examine the magnetic field evolution of the source region. The AIA images in seven extreme ultraviolet (EUV) wavelengths, with a pixel resolution of 0$\farcs$6 and a cadence of 12 s were used to analyse the coronal response. The event was also scanned in narrowband slit-jaw images (SJIs) from Interface Region Imaging Spectrograph (IRIS: de Pontieu et al. 2014); the time cadence and spatial resolution of each SJI are 36 seconds and 0$\farcs$332, respectively.

The observations from the different instruments were co-aligned by distinguishing an obvious identical feature (e.g., pores and sunspots) by naked eyes and cross correlations of the temporally closest images. We also investigated the emission properties of the flux rope with the differential emission measure (DEM) method by employing the sparse inversion code (Cheung et al. 2015; Su et al. 2018). In DEM method, the EM maps at different temperature ranges are obtained by a set of AIA images in six channels (i.e., 94, 131, 171, 193, 211, and 335~{\AA}).
%%%%%%%%%%%%%%%%%%%%%%%%%%%%%%%%%%%%%%%%%%%%%%%%%%%%%%%%%%%%%%%%%%%%%%%%%%%%%%%%%%%%%%%%%%%%%%%%%%%%%%%%%%%%%%%%%%%%%
\begin{figure*}
\centering
\includegraphics[width=8cm]{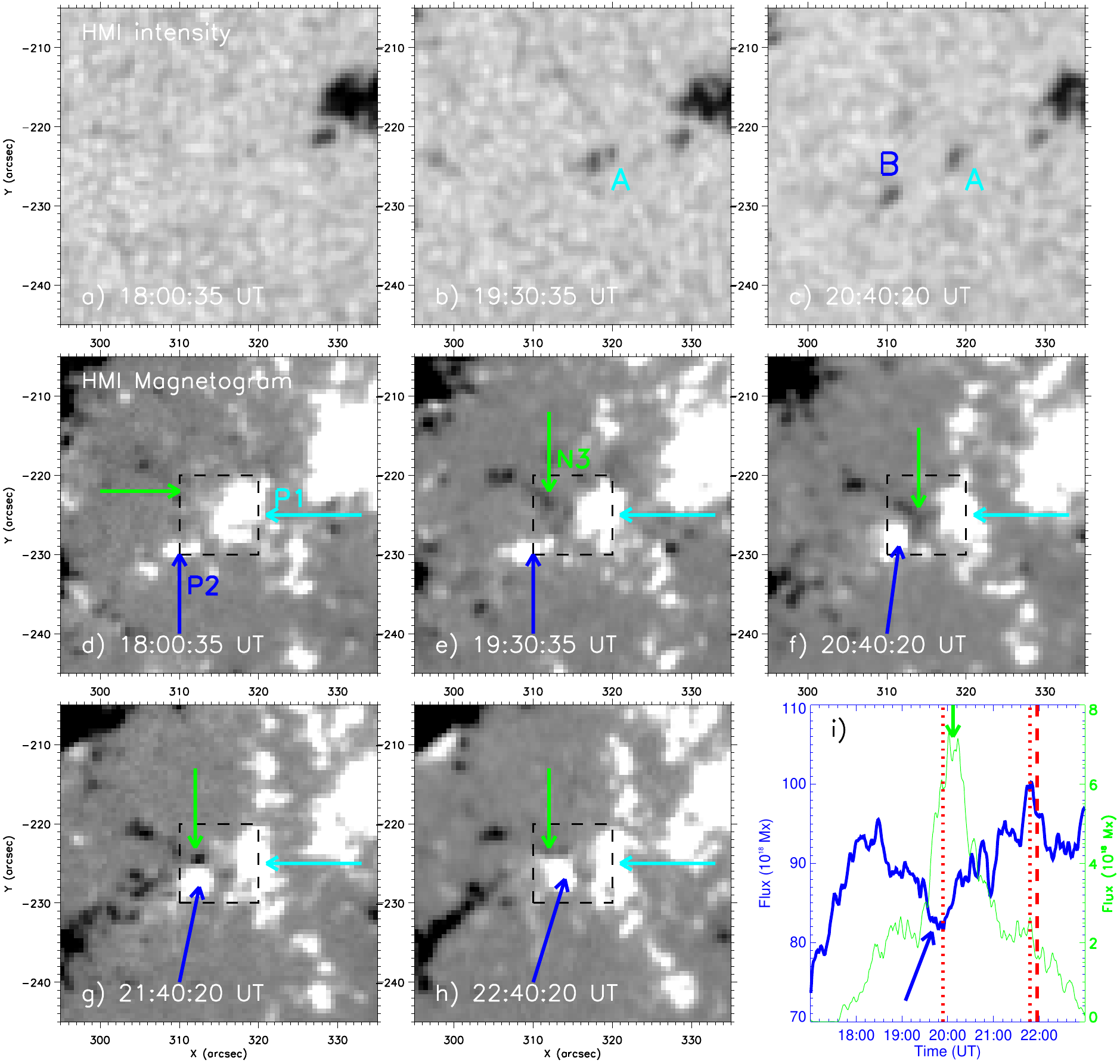}
\caption{Emergence and cancellation of magnetic flux in HMI intensity images and magnetograms. A and B represent the emerging pores, and the cyan, blue, and green arrows show the emerging magnetic polarities (P1, P2, and N3). The dashed box indicates the magnetic interaction region for P1-P2 and N3, and the positive (blue) and native (green) magnetic flux changes in this region is shown in panel (i). The blue and green arrows indicate the trough of the positive flux curve and the peak of the negative flux curve, respectively. The dotted lines and two dashed lines represent the timings of two brightenings and the eruption onset, respectively. The temporal evolution of magnetic fields and pores is shown in the online movie of Animation 1.}
\label{f2}
\end{figure*}
%%%%%%%%%%%%%%%%%%%%%%%%%%%%%%%%%%%%%%%%%%%%%%%%%%%%%%%%%%%%%%%%%%%%%%%%%%%%%%%%%%%%%%%%%%%%%%%%%%%%%%%%%%%%%%%%%%%%%
\section{Results}

\textbf{\subsection{Magnetic flux emergence, convergence, and cancellation}}

Figure 1 shows the overview of the source region of AR 12719 from the photosphere to the corona before the eruption ($\sim$18:04 UT). The AR 12719 consisted of a predominant positive sunspot and a minor negative sunspot (panels (a) and (b)) and was filled with clusters of coronal loops (panel (c)). The predominant positive sunspot was surrounded by a series of negative magnetic polarities, and this study concerns two negative polarities (N1-N2) in the east and two minor positive polarities (P1-P2) in the southeast of the predominant sunspot (panel (d)). Because of the magnetic interaction between N2 and the nearby positive polarities, some brightenings appeared before the eruption (panels (d) and (e)). However, this interaction region of N2 is not related to this study, and we primarily focus on the magnetic activities around P1 and the overlying loops connecting P1 and N1 at different heights (L1; panels (e) and (f)). In the GST images, H$\alpha$ superposed with contours of HMI longitudinal magnetic fields (panel (i)), bundles of filament threads rooting at P1 were related to the overlying L1. In GST the TiO image (panel (h)), the details of the predominant sunspot were very clear, but P1 and P2 were as yet invisible, comparing to those in panel (g).
%%%%%%%%%%%%%%%%%%%%%%%%%%%%%%%%%%%%%%%%%%%%%%%%%%%%%%%%%%%%%%%%%%%%%%%%%%%%%%%%%%%%%%%%%%%%%%%%%%%%%%%%%%%%%%%%%%%%%
\begin{figure*}
\centering
\includegraphics[width=8cm]{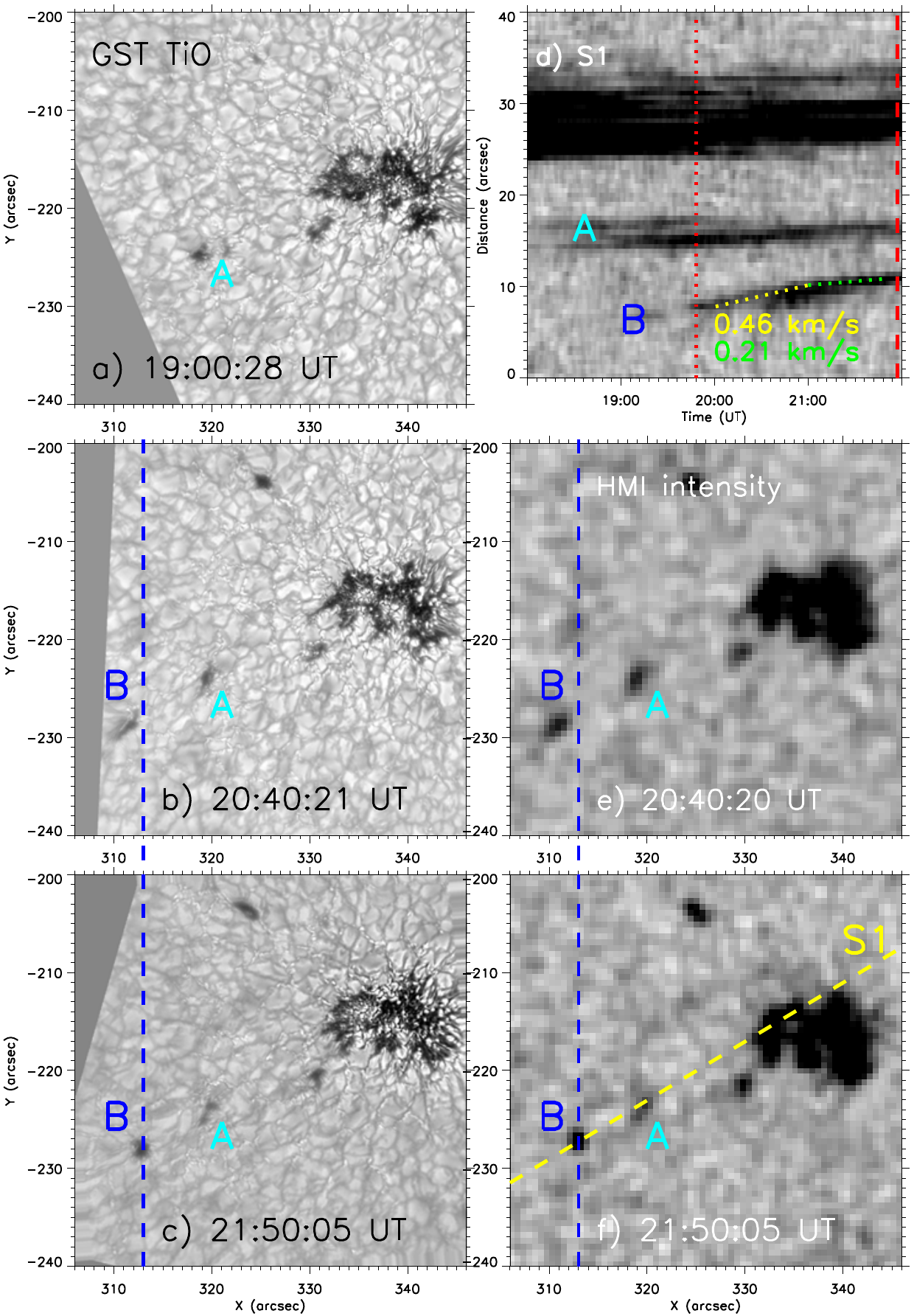}
\caption{Emergence and convergence of pores A and B in GST TiO and HMI intensity images. The blue dashed lines are reference lines to show the convergence of pores. The time-distance plot (panel (d)) of HMI intensity images along the yellow dashed line (S1 in (f)). The yellow and green dotted lines are used to derive the attached convergence speeds, and the red dotted and dashed lines represent the start of convergence of Pore B and the eruption onset, respectively.}
\label{f3}
\end{figure*}
%%%%%%%%%%%%%%%%%%%%%%%%%%%%%%%%%%%%%%%%%%%%%%%%%%%%%%%%%%%%%%%%%%%%%%%%%%%%%%%%%%%%%%%%%%%%%%%%%%%%%%%%%%%%%%%%%%%%%

In a few hours, two pores (A and B) successively appeared in the east of the predominant sunspot (panels (a)-(c) of Figure 2), as a result of the continuous magnetic flux emergence of P1-P2. During the emergence of P1-P2, some minor negative polarities (N3) emerged, approached P1-P2, and finally vanished (panels (d)-(h) of Figure 2). It is obvious that the emerging N3 first collided with P1 and then interacted with P2. The magnetic flux evolution (between 17:00 and 23:00 UT) in the magnetic interaction region for P1-P2 and N3 is shown in panel (i) of Figure 2. The negative flux has a fast increase and rapid decrease, with a net increase of $\sim$0. The positive flux of P1-P2 has a net increase of $\sim24 \times 10^{18} Mx$ in $\sim$6 hours, and shows a clear trough at $\sim$19:54 UT, which is closely associated with the peak of negative flux. It is evident that the continuous magnetic emergence and cancellation in the region is closely associated with the onset of the following eruption. On the other hand, P2 was moving toward P1 during its emergence, which is demonstrated by its location change to the bottom left corner of the black box.
%%%%%%%%%%%%%%%%%%%%%%%%%%%%%%%%%%%%%%%%%%%%%%%%%%%%%%%%%%%%%%%%%%%%%%%%%%%%%%%%%%%%%%%%%%%%%%%%%%%%%%%%%%%%%%%%%%%%%
\begin{figure*}
\centering
\includegraphics[width=16cm]{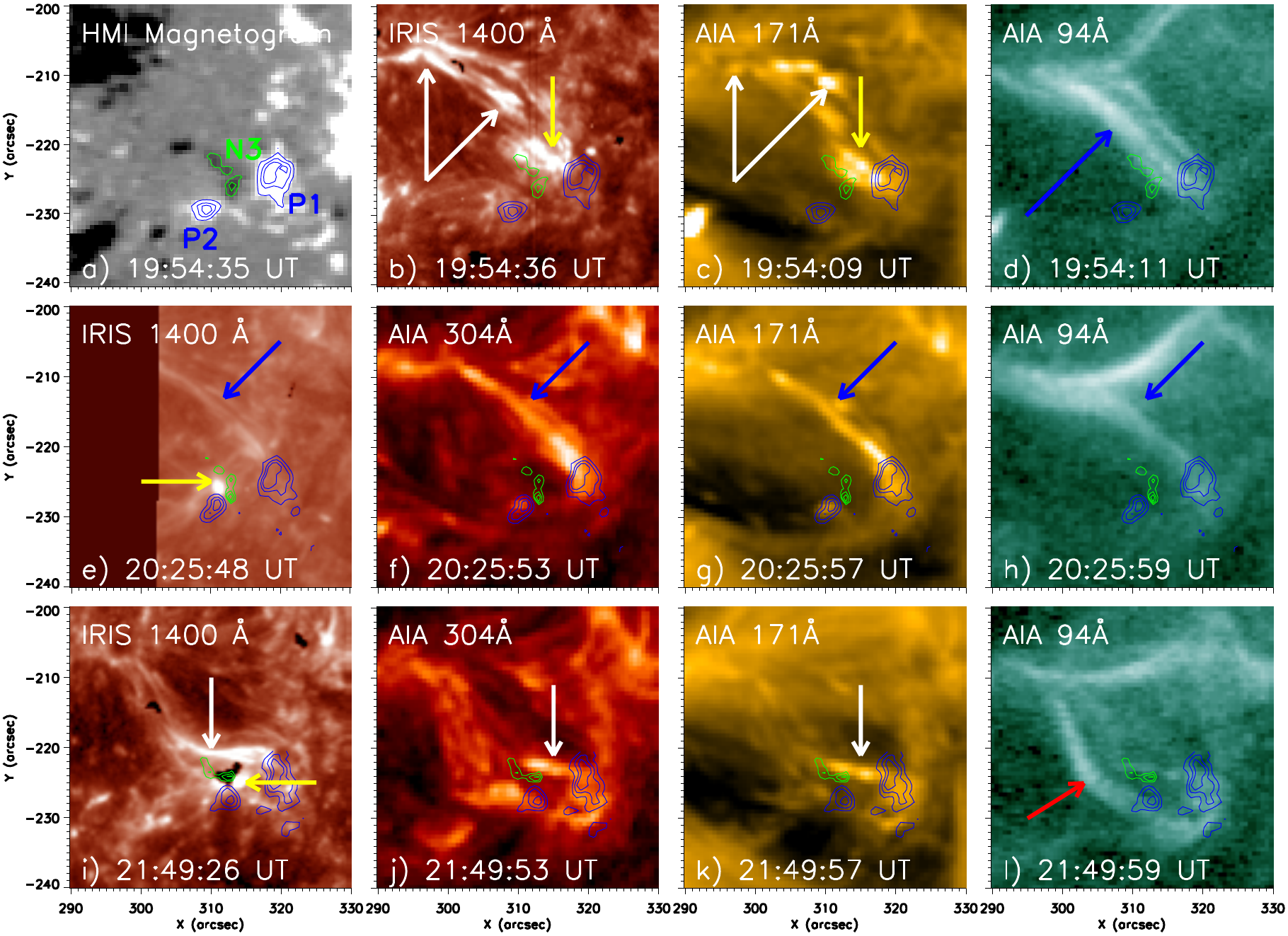}
\caption{Formation of the tiny flux rope in HMI magnetogram, IRIS 1400~{\AA}, and AIA 304, 171 and 94~{\AA}. The yellow and white arrows indicate the brightenings. The blue and red arrows point out the L1 and flux rope, respectively. Contours of HMI longitudinal magnetic fields at the closest time are superposed on all panels with positive (negative) fields in blue (green), and the levels for positive (negative) fields are 200 (50), 400 (100), and 600 (150) gauss. The formation of the tiny flux rope is shown in the online movie of Animation 2.}
\label{f4}
\end{figure*}
%%%%%%%%%%%%%%%%%%%%%%%%%%%%%%%%%%%%%%%%%%%%%%%%%%%%%%%%%%%%%%%%%%%%%%%%%%%%%%%%%%%%%%%%%%%%%%%%%%%%%%%%%%%%%%%%%%%%%

The displacement of P2 appeared as the convergence of Pore B in images of GST TiO and HMI intensity (Figure 3). During the convergence, Pore B became strong, and Pore A decayed. The emergence and convergence of pores is clearly shown in the time-distance plot (panel (d)) of HMI intensity images along the selected path (S1; panel (f)). The beginning of the emergence of Pore B was about one hour later than the appearance of Pore A, and was two hours before the eruption onset. Then, Pore B moved toward Pore A at an initial speed of $\sim$0.46 km s$^{-1}$, and slowly decreased to a speed of $\sim$0.21 km s$^{-1}$.
%%%%%%%%%%%%%%%%%%%%%%%%%%%%%%%%%%%%%%%%%%%%%%%%%%%%%%%%%%%%%%%%%%%%%%%%%%%%%%%%%%%%%%%%%%%%%%%%%%%%%%%%%%%%%%%%%%%%%
\begin{figure*}
\centering
\includegraphics[width=16cm]{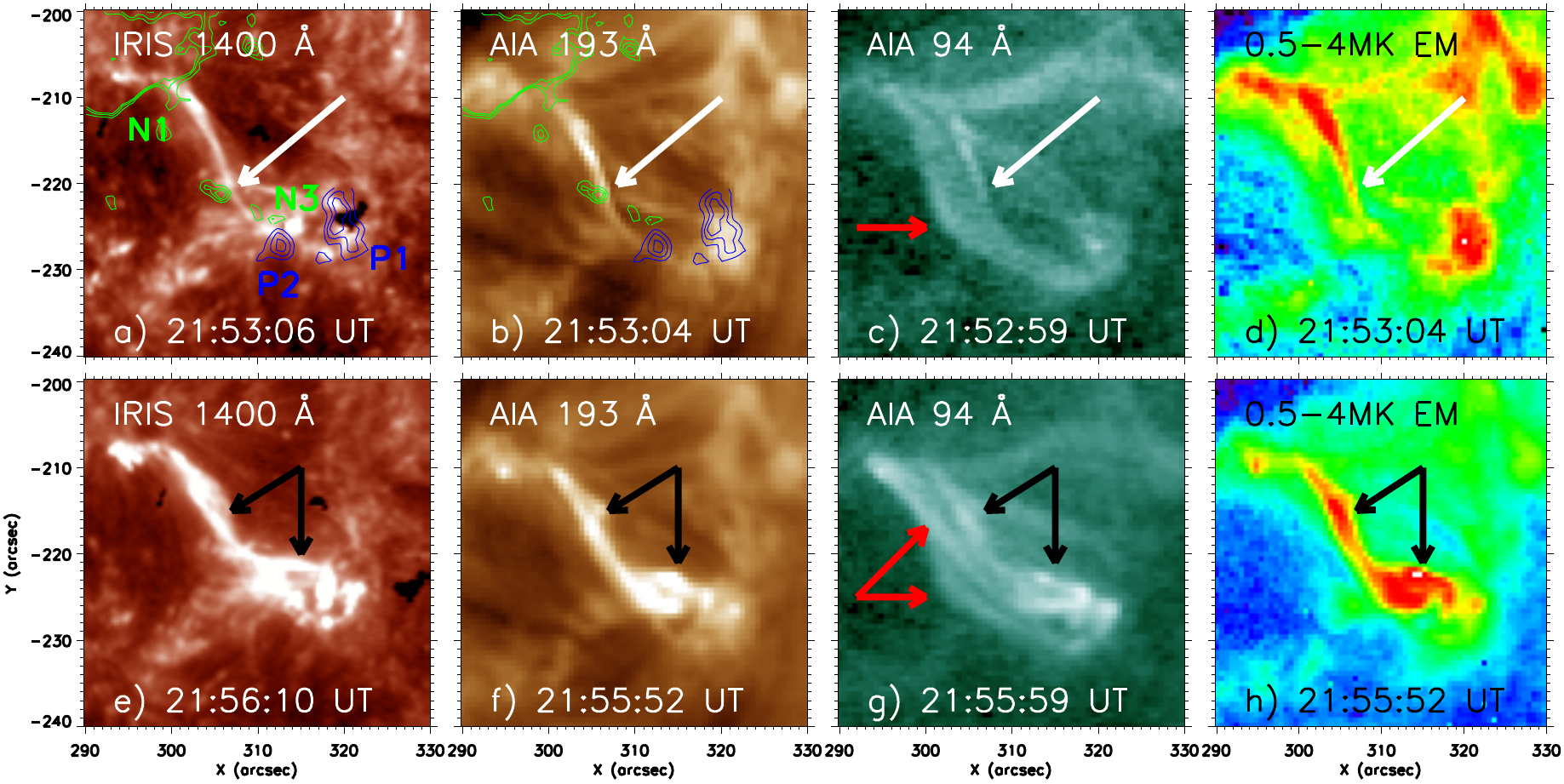}
\caption{Formation of the sigmoid in IRIS 1400~{\AA}, AIA 193 and 94~{\AA}, and EM maps at temperature range of 0.5-4 MK. The white and red arrows indicate the sigmoid and the flux rope, respectively. The black arrows show the brightenings. The contours of HMI longitudinal magnetic fields at the closest time are superposed on all panels with positive (negative) fields in blue (green), and the levels for positive (negative) fields are 200 (50), 400 (100), and 600 (150) gauss.}
\label{f5}
\end{figure*}
%%%%%%%%%%%%%%%%%%%%%%%%%%%%%%%%%%%%%%%%%%%%%%%%%%%%%%%%%%%%%%%%%%%%%%%%%%%%%%%%%%%%%%%%%%%%%%%%%%%%%%%%%%%%%%%%%%%%%

\textbf{\subsection{Formation and eruption of flux rope}}
During the period of the continuous magnetic activities in the interaction region of P1-P2 and N3, brightenings and loop transformations in the upper atmosphere appeared (Figure 4). At $\sim$19:54 UT (top panels), the brightenings occurred at the southern end of the overlying L1, which were associated with the underlying magnetic cancellation between N3 and P1 (magnetic field contours)), and some other brightenings also appeared at the center and at the northern end of L1. Interestingly, L1 became sheared, which was first seen in AIA 94~{\AA} (panel (d)). In the following half hour, L1 became very sheared, which was simultaneously seen in IRIS UV and AIA EUV channels (panels (e)-(h)). We note that a bright point appeared between P2 and N3 and moved westward as the convergence motion of P1 (panels (e) and (i)). Following the motion of the bright point, at $\sim$21:49 UT (bottom panels), some brightenings appeared again at the southern portion of L1. On the other hand, the timings of two major brightenings around L1 at $\sim$19:54 UT and $\sim$21:49 UT were indicated by the red dotted lines in Figure 2(i). It is evident that the brightenings in the upper atmosphere have a close temporal and spatial relationship with the continuous magnetic flux emergence, convergence, and cancellation in the photosphere.
%%%%%%%%%%%%%%%%%%%%%%%%%%%%%%%%%%%%%%%%%%%%%%%%%%%%%%%%%%%%%%%%%%%%%%%%%%%%%%%%%%%%%%%%%%%%%%%%%%%%%%%%%%%%%%%%%%%%%
\begin{figure*}
\centering
\includegraphics[width=16cm]{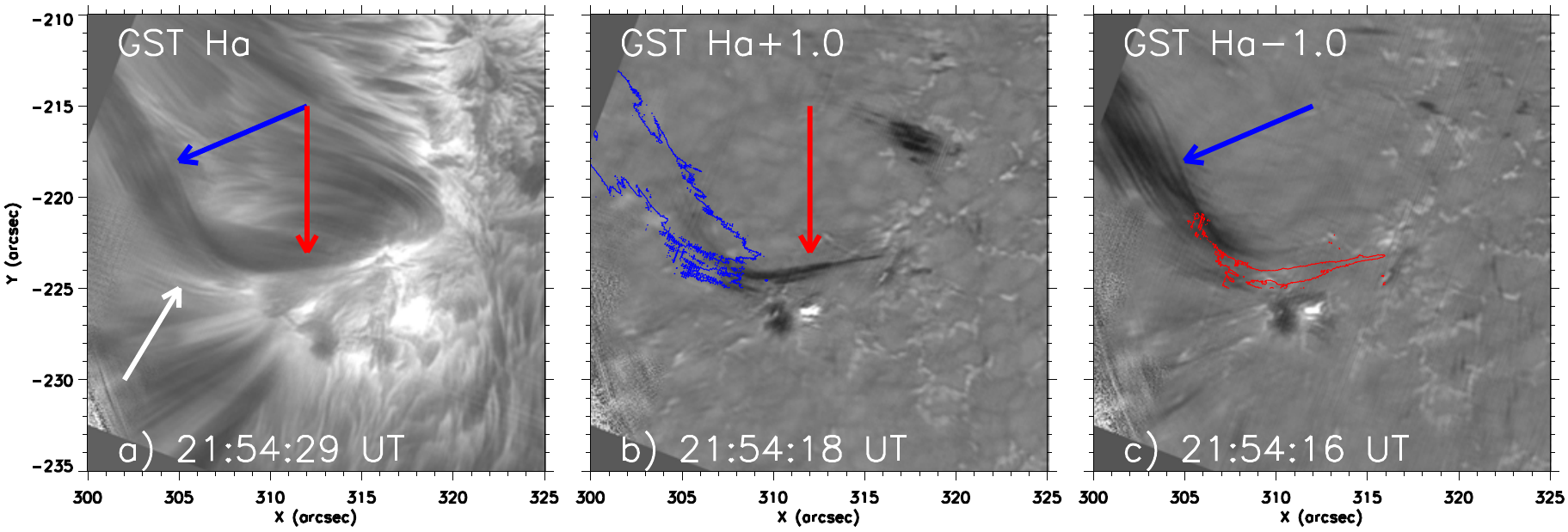}
\caption{Twisted filament threads (the white arrow) in GST H$\alpha$ center and wings. The red and blue arrows indicate the parts with blueshift and redshift signatures, respectively. The contours of filament portions with blueshift (blue) and redshift (red) signatures are superposed in panels (b)-(c).}
\label{f6}
\end{figure*}
%%%%%%%%%%%%%%%%%%%%%%%%%%%%%%%%%%%%%%%%%%%%%%%%%%%%%%%%%%%%%%%%%%%%%%%%%%%%%%%%%%%%%%%%%%%%%%%%%%%%%%%%%%%%%%%%%%%%%

Some minutes after the formation of the tiny flux rope, an inverse S-shaped transient sigmoid occurred, and was clearly seen in images of IRIS UV and AIA EUV,  and the sigmoid structure also appeared in the EM map at a temperature range of 0.5-4 MK (Figure 5). The transient sigmoid lay along the polarity inversion line (PIL) of P1-P2 and N1-N3 and likely represents the axis of the overlying flux rope (Rust \& Kumar 1996). It is possible that the transient sigmoid was under the newly-formed twisted flux rope (the red arrow in panel (c)). Shortly after the appearance, the sigmoid and overlying flux rope erupted together, and left brightenings behind (panels (e)-(h) in Figure 5). This is consistent with the fact that transient sigmoids always occur as the flux rope loses equilibrium and erupts (Gibson et al. 2006).  In addition, the southern part of the filament threads beneath also became twisted structures (Figure 6) in H$\alpha$ center, and showed the simultaneous blueshift and redshift signatures in H$\alpha$ blue and red wing images, respectively. The formation of twisted filament threads corresponded to the formation of the twisted flux rope that contained a transient sigmoid.
%%%%%%%%%%%%%%%%%%%%%%%%%%%%%%%%%%%%%%%%%%%%%%%%%%%%%%%%%%%%%%%%%%%%%%%%%%%%%%%%%%%%%%%%%%%%%%%%%%%%%%%%%%%%%%%%%%%%%
\begin{figure*}
\centering
\includegraphics[width=8cm]{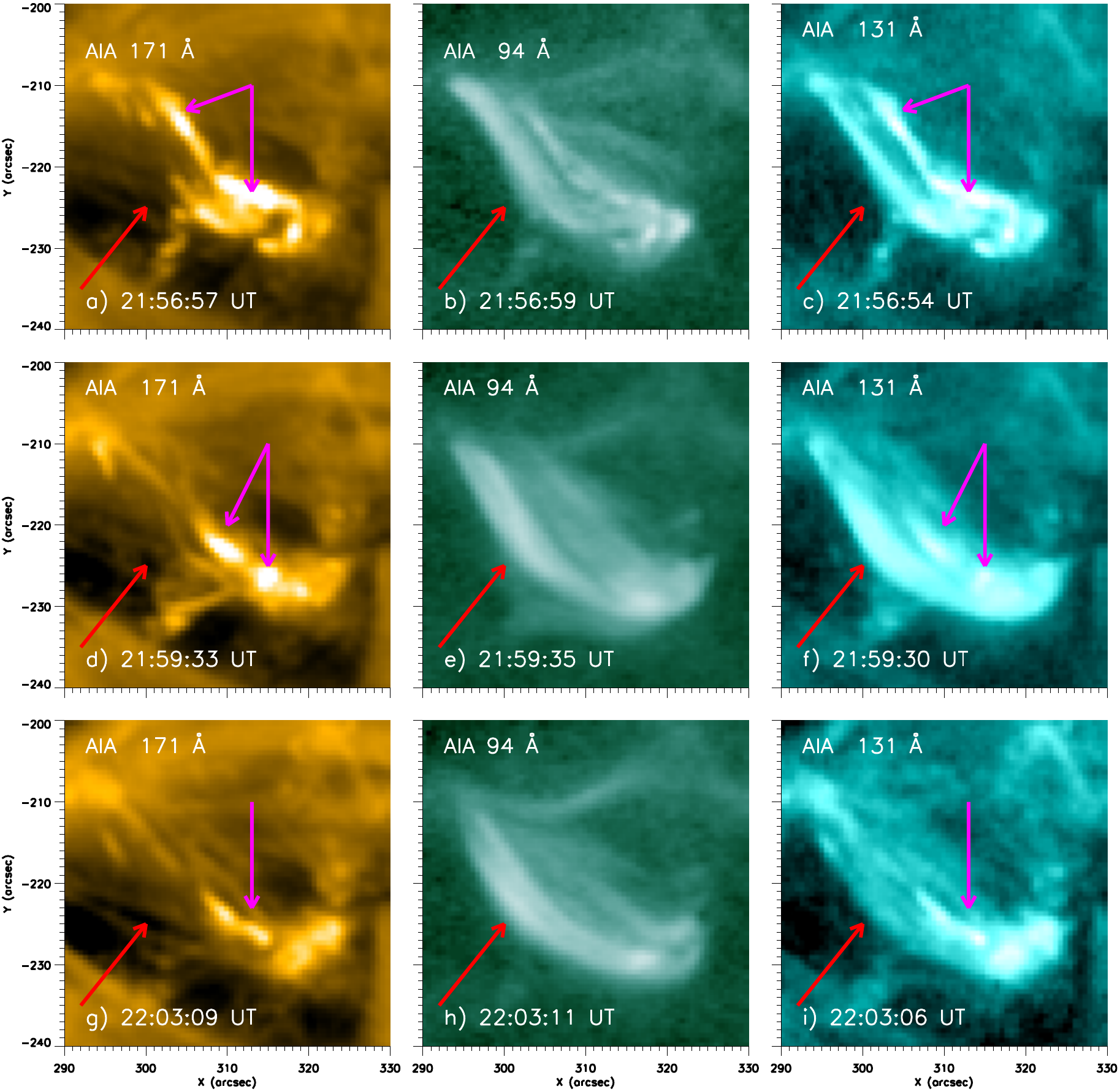}
\caption{Eruption of the tiny flux rope in AIA 171, 131 and 94~{\AA}. The red arrows indicate the erupting flux rope, and the pink arrows show the flare ribbons. The eruption of the tiny flux rope is shown in the online movie of Animation 3.}
\label{f7}
\end{figure*}
%%%%%%%%%%%%%%%%%%%%%%%%%%%%%%%%%%%%%%%%%%%%%%%%%%%%%%%%%%%%%%%%%%%%%%%%%%%%%%%%%%%%%%%%%%%%%%%%%%%%%%%%%%%%%%%%%%%%%

The evolution of eruption of the flux rope is shown in Figure 7. The rising flux rope disappeared in AIA cool channel of 171~{\AA} channels, but remained visible in AIA hot channels of 131 and 94~{\AA}. This is likely because the rising flux rope was heated to a much higher temperature. During the eruption, the twisted flux rope transformed into a smooth shape (panels (e) and (f)). On the other hand, the eruption initiated a B1.6 flare, but the flare ribbons quickly faded away in a few minutes (left and right panels). The rising flux rope also became weak in the hot AIA 94 and 131~{\AA} (panels (h)-(i)). Hence, the eruption was failed, which was possibly restricted by the overlying bundles of AR coronal loops (Figure 1(c)).

%{\bf The transformation, heating, and cooling of the erupting flux rope} is also shown in EM maps at different temperature ranges (Figure 8). In the EM maps at the range of 4-10 MK (left panels), the rising flux rope (red arrows) showed clearly the transformation to the smooth shape, and the lower bright structures possibly coincided with the post-eruption loops (black arrows). In the EM maps at the range of 10-30 MK (right panels), it is very obvious for the process of heating and cooling of the rising flux rope (red arrows) and for the fading of flare ribbons (pink arrows) after the eruption.

%%%%%%%%%%%%%%%%%%%%%%%%%%%%%%%%%%%%%%%%%%%%%%%%%%%%%%%%%%%%%%%%%%%%%%%%%%%%%%%%%%%%%%%%%%%%%%%%%%%%%%%%%%%%%%%%%%%%%
%\begin{figure*}
%\centering
%\includegraphics[width=8cm]{fig8.pdf}
%\caption{The eruption of the tiny flux rope in EM maps at temperature ranges of 4-10 MK and 10-30 MK. The red and pink arrows indicate the flux rope and the flare ribbons, respectively. The black arrows show the post-eruption loops.}
%\label{f8}
%\end{figure*}
%%%%%%%%%%%%%%%%%%%%%%%%%%%%%%%%%%%%%%%%%%%%%%%%%%%%%%%%%%%%%%%%%%%%%%%%%%%%%%%%%%%%%%%%%%%%%%%%%%%%%%%%%%%%%%%%%%%%%

\section{Discussion and conclusions}

In this paper, we present the formation and eruption of a tiny flux rope in the center of AR 12719, using high-quality multiwavelength observations from GST, HMI, AIA, and IRIS. With aim to study the physical mechanisms of the formation and eruption of a tiny flux rope, we investigated the observational characteristics occurring simultaneously in different layers of solar atmosphere, from the magnetic activities in the bottom photosphere to the upper atmospheric responds extending to the higher corona.

In the bottom photosphere, there were continual magnetic activities in the source region through the event. Two positive magnetic polarities of P1 and P2 continuously emerged, which led to the successive emergence of two pores of A and B (Figure 2). Interestingly, P2 and Pore B kept moving toward P1 and Pore A, respectively, which indicates the existence of a magnetic flux convergence (Figure 3). Meanwhile, P1-P2 successively interacted with the ambient emerging negative magnetic polarity (N3) (Figure 2). In the upper atmosphere, during the consecutive magnetic activities, the intermittent brightenings occurred around the overlying loops (L1), and L1 first became sheared and eventually evolved into a tiny twisted flux rope (Figure 4). Shortly after the formation of the tiny flux rope, a transient inverse S-shaped sigmoid appeared along the PIL of P1-P2 and N1-N3 in the interior of the flux rope, and then quickly erupted with the newly formed flux rope (Figures 5 and 6). The twisted flux rope was transformed into a smooth shape during the eruption. However, the flux rope stopped rising quickly, and then soon cooled down (Figure 7), which revealed a failed eruption. The time sequence of the main observational activities for this event is shown in Figure 8.
%%%%%%%%%%%%%%%%%%%%%%%%%%%%%%%%%%%%%%%%%%%%%%%%%%%%%%%%%%%%%%%%%%%%%%%%%%%%%%%%%%%%%%%%%%%%%%%%%%%%%%%%%%%%%%%%%%%%%
\begin{figure*}
\centering
\includegraphics[width=12cm]{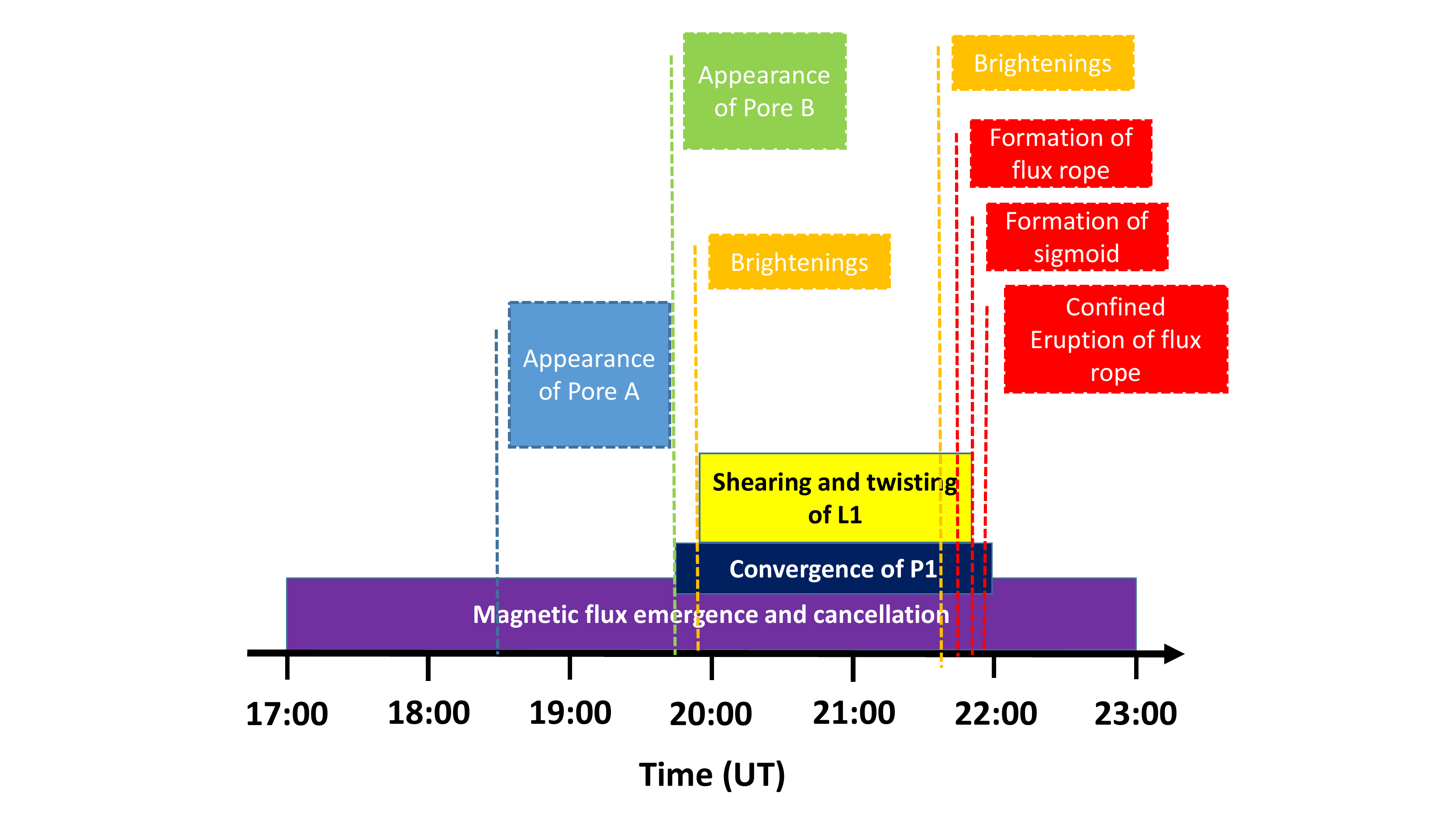}
\caption{Timeline of the main observational activities for this event.}
\label{f8}
\end{figure*}
%%%%%%%%%%%%%%%%%%%%%%%%%%%%%%%%%%%%%%%%%%%%%%%%%%%%%%%%%%%%%%%%%%%%%%%%%%%%%%%%%%%%%%%%%%%%%%%%%%%%%%%%%%%%%%%%%%%%%

How do we identify that the coronal loops (L1) in the AR center evolved into a flux rope? It is noteworthy that, after becoming very sheared, the apex of L1 shifted to the left side viewing along its axis direction (from P1 to N1) of magnetic field as a twisted structure (Figure 4(l)). The appearance of the inverse S-shaped sigmoid also indicates the kinked axis of a flux rope (Figure 5). Furthermore, the filament threads beneath L1 simultaneously showed the blueshift and redshift signatures (Figure 6), which likely implies the formation of twisted structures in the chromosphere. In addition, the erupting structure disappeared in AIA 171~{\AA}, but became bright in AIA 131 and 94~{\AA} at the beginning of the eruption. This possibly indicates the existence of a hot channel as the manifestation of the flux rope (Figure 7). Hence, we suggest that a tiny twisted flux rope formed by the evolution from L1 in the AR center.

In general, flux ropes can be formed by the bodily emergence from the convection zone, or as a result of photospheric motions. In this paper, the observational results exclude the possibility of the bodily emergence. The transformation of L1 and around intermittent brightenings (Figure 4) had a close temporal and spatial relationship with the photospheric magnetic flux emergence, convergence, and cancellation (Figures 2 and 3). It has been reported that the magnetic flux convergence and cancellation can play critical role in the formation of flux ropes (van Ballegooijen \& Martens 1989). However, the magnetic flux convergence in this study is of the same polarities (P1 and P2), which is different from the convergence of opposite polarities in previous cases. But, the convergence of P2 toward P1 led to the magnetic flux cancellation between P2 and N3, which has the same consequence as the convergence of opposite polarities does. Based on the above results and discussions, we propose a scenario for the formation of the tiny flux rope in the schematic diagram of Figure 9. The positive polarities of P1-P2 and the negative polarity of N3 rapidly emerge beneath L1, which connected with P1 and N1, and P2 also moves toward P1 during the emergence. As a consequence of the close location relations, N3 sequentially canceled with P1 and the converging P2 quickly. As the result of successive magnetic cancellation, the overlying L1 first became sheared and finally formed a twisted flux rope. Hence, the formation of the tiny flux rope is intimately associated with the magnetic flux emergence, convergence, and cancellation in the photosphere.
%%%%%%%%%%%%%%%%%%%%%%%%%%%%%%%%%%%%%%%%%%%%%%%%%%%%%%%%%%%%%%%%%%%%%%%%%%%%%%%%%%%%%%%%%%%%%%%%%%%%%%%%%%%%%%%%%%%%%
\begin{figure*}
\centering
\includegraphics[width=8cm]{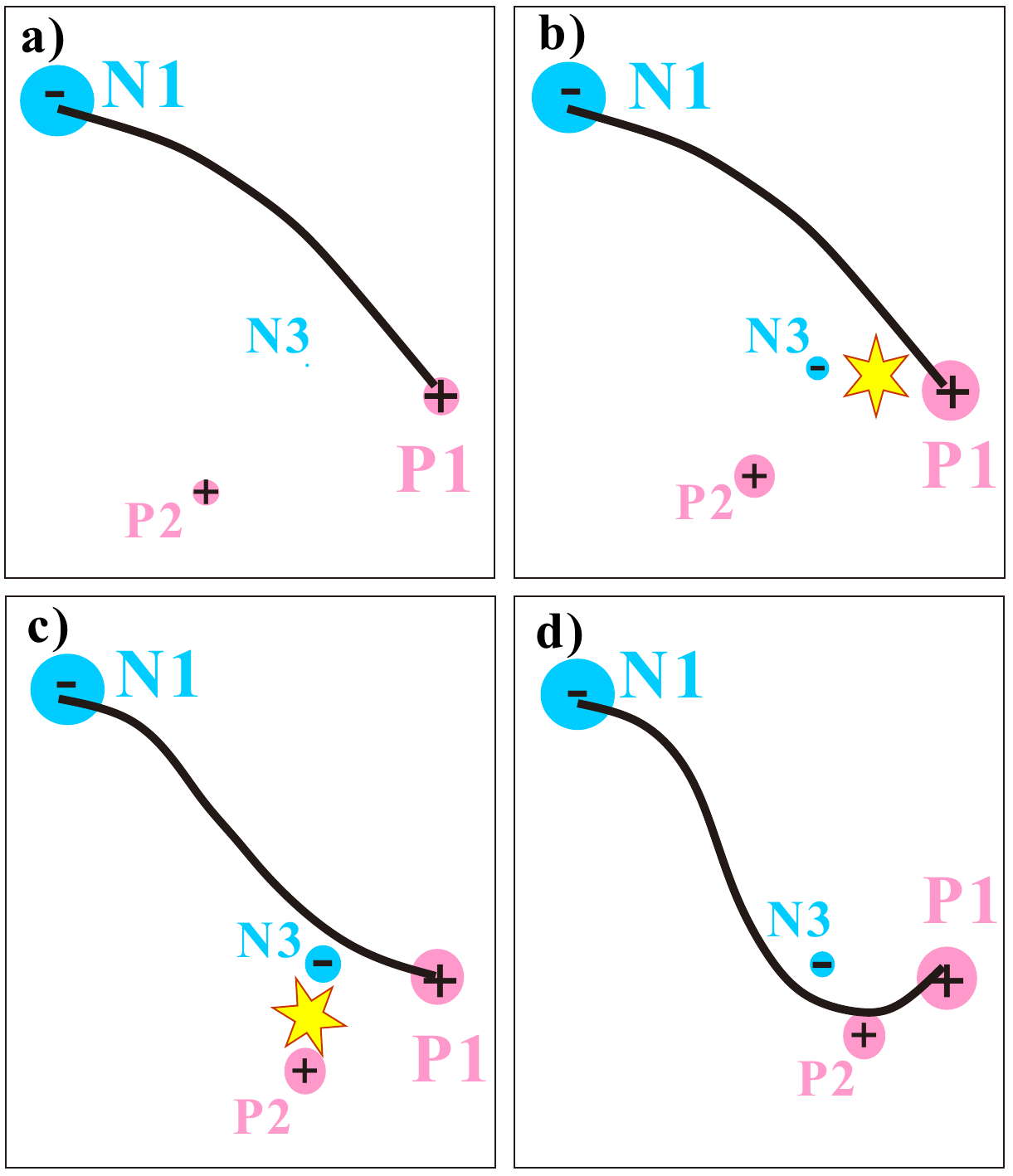}
\caption{Schematic showing the scenario of the formation of the tiny flux rope (black line) associated with the magnetic flux emergence, convergence, and cancellation of P1-P2 and N3. The yellow sparks indicate the magnetic cancellation sites.}
\label{f9}
\end{figure*}

The magnetic flux emergence, convergence, and cancellation lasted through the eruption {\bf process} (panel (i) of Figure 2 and panel (d) of Figure 3), after the formation of the tiny flux rope. On the other hand, the appearance of a transient sigmoid (Figure 5) can be an indicator of the flux rope undergoing the kink instability (Fan \& Gibson 2004; Kliem et al. 2004; Gibson et al. 2006). Therefore, the initiation of the tiny flux rope eruption should involve several important factors. The strength of the constraining fields is very important for the destination of an eruption (Liu et al. 2016), and the strong constraining fields can act as a confining cage (Ji et al. 2003; Cheng et al. 2015; Amari et al. 2018; Yang et al. 2018; Zheng et al. 2019). In this work, the tiny flux rope located at the AR center was covered by clusters of AR loops (Figure 1). Eventually, the erupting flux rope halted and cooled down in a few minutes, and only led to a confined small flare at the base (Figure 7). Hence, the failed eruption of the tiny flux rope was possibly due to the strong restriction of overlying AR loops.

In conclusion, it is believed that a tiny flux rope formed in the AR center, and its formation and helicity were closely associated with the continuous magnetic flux emergence, convergence, and cancellation in the photosphere. The eruption of the tiny flux rope took place during a period of magnetic flux emergence, convergence, and cancellation, and it is also possible that the kink instability may have occurred during this period. We suggest that magnetic flux emergence, convergence, and cancellation are very important for the formation and eruption of the tiny flux rope. Further and better observations should be investigated in more detail to understand the formation and eruption of flux ropes.

\begin{acknowledgements}
The authors thank the anonymous referee for constructive comments and thoughtful suggestions. SDO is a mission of NASA's Living With a Star Program. BBSO operation is supported by NJIT and US NSF AGS-1821294 grant. GST operation is partly supported by the Korea Astronomy and Space Science Institute, the Seoul National University, and the Key Laboratory of Solar Activities of Chinese Academy of Sciences (CAS) and the Operation, Maintenance and Upgrading Fund of CAS for Astronomical Telescopes and Facility Instruments. We gratefully acknowledge the usage of data from the SDO and IRIS spacecrafts, and the ground-based GST project. This work is supported by grants NSFC 11790303 and U1731101, and Young Scholars Program of Shandong University, Weihai, 2016WHWLJH07. H. Q. Song is supported by the Natural Science Foundation of Shandong Province JQ201710.

\end{acknowledgements}


\begin{thebibliography}{}
\bibitem{amari18} Amari, T., Canou, A., Aly, J.-J., Delyon, F., \& Alauzet, F. 2018, Nature, 554, 211

\bibitem{antiochos99} Antiochos, S. K., DeVore, C. R. \& Klimchuk, J. A. 1999, \apj, 510, 485

\bibitem{aula10} Aulanier, G., T{\"o}r{\"o}k, T., D{\'e}moulin, P., \& DeLuca, E. E. 2010, \apj, 708, 314

\bibitem{cao10} Cao, W., Gorceix, N., Coulter, R., Ahn, K., Rimmele, T. R. \& Goode, P. R. 2010, Astronomische Nachrichten, 331, 636

\bibitem{chen00} Chen, P. F. \& Shibata, K. 2000, \apj, 545, 524

\bibitem{cheng17} Cheng, X., Guo, Y. \& Ding, M. D. 2017, Science China Earth Sciences, 60, 1383

\bibitem{cheng15} Cheng, X., Hao, Q., Ding, M. D., Liu, K., Chen, P. F., Fang, C., \& Liu, Y. D. 2015, \apj, 809, 46

\bibitem{cheng11} Cheng, X., Zhang, J., Liu, Y. \& Ding, M. D. 2011, \apjl, 732, L25

\bibitem{cheung15} Cheung, M. C. M., Boerner, P., Schrijver, C. J., et al. 2015, \apj, 807, 143


\bibitem{depon14} de Pontieu, B., Title, A. M., Lemen, J. R., et al. 2014, \solphys, 289, 2733


\bibitem{fan01} Fan, Y. 2001, \apjl, 554, L111

\bibitem{fan09} Fan, Y. 2009, \apj, 697, 1529

\bibitem{fan04} Fan, Y., \& Gibson, S. E. 2004, \apj, 609, 1123

\bibitem{forbes91} Forbes, T. G. \& Isenberg, P. A. 1991, \apj, 373, 294

\bibitem{gibson06} Gibson, S. E., Fan, Y., T{\"o}r{\"o}k, T., \& Kliem, B. 2006, \ssr, 124, 131

\bibitem{green11} Green, L. M., Kliem, B. \& Wallace, A. J. 2011, \aap, 526, A2

\bibitem{ji03} Ji, H., Wang, H., Schmahl, E. J., Moon, Y.-J., \& Jiang, Y. 2003, \apjl, 595, L135

\bibitem{kliem04} Kliem, B., Titov, V. S., \& T{\"o}r{\"o}k, T. 2004, \aap, 413, L23

\bibitem{kliem06} Kliem, B., \& T{\"o}r{\"o}k, T. 2006, Physical Review Letters, 96, 255002

\bibitem{kumar15} Kumar, P., Yurchyshyn, V., Wang, H. \& Cho, K. S. 2015, \apj, 809, 83


\bibitem{leak13} Leake, J. E, Linton, M. G. \& T{\"o}r{\"o}k, T. 2013, \apj, 778, 99

\bibitem{lemen12} Lemen, James R., Title, Alan M., Akin, David J., Boerner, Paul F. et al. 2012, \solphys, 275, 17

\bibitem{lin00} Lin, J., \& Forbes, T. G. 2000, Journal of Geophysical Research, 105, 2375

\bibitem{lites95} Lites, B. W., Low, B. C., Martinez Pillet, V., Seagraves, P., Skumanich, A., Frank, Z. A., Shine, R. A., Tsuneta, S. 1995, \apj, 446, 877

\bibitem{liu18} Liu, L. J, Cheng, X., Wang, Y. M., Zhou, Z. J., Guo, Y. \& Cui, J. 2018, \apjl, 867, L5

\bibitem{liu16} Liu, R., Kliem, B., Titov, V. S., et al. 2016, \apj, 818, 148

\bibitem{liu08} Liu Y. 2008, \apjl, 679, L151

\bibitem{manchester04} Manchester, W., IV, Gombosi, T., DeZeeuw, D., Fan, Y. 2004, \apj, 610, 588

\bibitem{martin92} Martin S. F., Marquette W. H. and Bilimoria R. 1992 ASP Conf. Ser. 27, The Solar Cycle Pattern in the Direction of the Magnetic Field along the Long Axes of Polar Filaments ed K. L. Harvey (San Francisco, CA: ASP) 53

\bibitem{moor01} Moore, R. L., Sterling, A. C., Hudson, H. S., \& Lemen, J. R. 2001, \apj, 532, 833

\bibitem{patsourakos13} Patsourakos, S., Vourlidas, A. \& Stenborg. G. 2013, \apj, 764, 125

\bibitem{pesnell12} Pesnell, W. Dean, Thompson, B. J., Chamberlin, P. C. et al. 2012, \solphys, 275, 3


%\bibitem{pevtsov03} Pevtsov A. A., Canfield, R. C. \& Latushko, S. M. 2001, \apjl, 549 L261

\bibitem{rust96} Rust D. M. \& Kumar A. 1996 \apjl, 464, L199

\bibitem{savcheva12} Savcheva, A. S., van Ballegooijen, A. A. \& DeLuca, E. E. 2012, \apj, 744, 78

\bibitem{sche12} Scherrer, P. H., Schou, J., Bush, R. I., Kosovichev, A. G., et al. 2012, \solphys, 275, 207

\bibitem{shiba95} Shibata, K., Masuda, S., Shimojo, M., et al. 1995, \apjl, 451, L83

\bibitem{song14} Song, H. Q., Zhang, J., Chen, Y. \& Cheng, X. 2014, \apjl, 792, L40

\bibitem{su18} Su, Y., Veronig, A. M., Hannah, I. G., et al. 2018, \apjl, 856, L17

\bibitem{sun15} Sun, X., Bobra, M. G., Hoeksema, J. T., Liu, Y., Li, Y., Shen, C., Couvidat, S., Norton,
A. A. \& Fisher, G. H. 2015, \apjl, 804, L28


\bibitem{threfall18} Threlfall, J., Hood, A. W. \& Priest, E. R., \solphys, 2018, 293, 98


\bibitem{torok04} T{\"o}r{\"o}k, T., Kliem B. \& Titov, V. S. 2004, \aap, 413, L27

\bibitem{vanball89} van Ballegooijen, A. A. \&, Martens, P. C. H. 1989, \apj, 343, 971


\bibitem{wang17} Wang, J., Liu, R., Wang, Y. M., Hu, Q., Shen, C. L., Jiang, C. W. \& Zhu, C. M. 2017, Nature Comm., 8, 1330

\bibitem{woger08} W{\"o}ger, F., von der L{\"u}he, O., Reardon, K. 2008, \aap, 488, 375


\bibitem{xia14} Xia, C., Keppens, R. \& Guo. Y. 2014, \apj, 780, 130


\bibitem{yan16} Yan, X. L., Priest, E. R., Guo, Q. L., Xue, Z. K., Wang, J. C. \& Yang, L, H. 2016, \apj, 832, 23

\bibitem{yang18} Yang, S. \& Zhang, J. 2018, \apjl, 860, L25

\bibitem{yang19} Yang, S., Zhang, J., Song, Q., Bi, Y. \& Li, T. 2019, \apj, 878, 38

%\bibitem{yeates09} Yeates, A. R. \& Mackay, D. H. 2009 \solphys, 254, 77

\bibitem{zhang12} Zhang, J., Cheng, X. \& Ding, M. D. 2012, Nature Comm., 3, 747

\bibitem{zhao17} Zhao, X. Z., Xia, C., Keppens, R. \& Gan, W. 2017, \apj, 841, 106

\bibitem{zheng19} Zheng, R. S., Yang, S. H., Rao, C. H., Liu, Y. Y., Zhong, L. B., Wang, B., Song, H. Q., Li, Z. \& Chen, Y. 2019, \apj, 875, 71

%\bibitem{zhou20} Zhou, Z., Liu, R., Cheng, X., Jiang, C., Wang, Y., Liu, L., Cui, J. 2020, \apj, 891, 180

\end{thebibliography}
\end{document}